\documentclass[aip, jcp, preprint]{revtex4-1}
\usepackage[utf8]{inputenc}

\usepackage[english]{babel}
\usepackage{amsmath}
\usepackage{amsthm}
\usepackage{amssymb}
\usepackage{graphicx}
\usepackage{physics}
\usepackage{array}
\usepackage{booktabs}
\usepackage{float}
\usepackage[toc,page]{appendix}
\restylefloat{table}
\usepackage[colorinlistoftodos]{todonotes}
\usepackage[colorlinks=true, allcolors=blue, urlcolor=black]{hyperref}
\usepackage[margin=1.25in]{geometry}
\usepackage{fancyhdr}

\usepackage{xfrac}
\usepackage{xcolor}

\newcommand{\eqnn}[2]{Eqs.~(\ref{#1}) and (\ref{#2})}

\begin{document}

\title{Testing the quasicentroid molecular dynamics method on gas-phase ammonia}

\author{Christopher~Haggard}
\affiliation{Yusuf Hamied Department of Chemistry, University of Cambridge, Lensfield Road, Cambridge,
CB2 1EW, UK.}
\author{Vijay Ganesh Sadhasivam}
\affiliation{Yusuf Hamied Department of Chemistry, University of Cambridge, Lensfield Road, Cambridge,
CB2 1EW, UK.}
\author{George~Trenins}
\affiliation{Laboratory of Physical Chemistry, ETH Z\"urich, 8093 Z\"urich, Switzerland.}
\author{\\Stuart C.~Althorpe}
\affiliation{Yusuf Hamied Department of Chemistry, University of Cambridge, Lensfield Road, Cambridge,
CB2 1EW, UK.}
\date{November 2021}

\begin{abstract} Quasicentroid molecular dynamics (QCMD) is a path-integral method for approximating nuclear quantum effects in dynamics simulations, which has given promising results for gas- and condensed-phase water.  Here, by simulating the infrared spectrum of gas-phase ammonia, we test the feasibility of extending QCMD beyond water. Overall, QCMD works as well for ammonia as for water, reducing or eliminating blue shifts from the classical spectrum without introducing the artificial red-shifts or broadening associated with other imaginary-time path-integral methods. However, QCMD gives only a modest improvement over the classical spectrum for the position of the symmetric bend mode, which is highly anharmonic (since it correlates with the inversion pathway). We expect QCMD to have similar problems with 
large-amplitude degrees of freedom in other molecules, but otherwise to work as well as for water.
\end{abstract}

\maketitle

\section{Introduction}

A variety of imaginary-time path-integral\cite{Chandler1981,Parrinello1984} methods, including (thermostatted) ring-polymer molecular dynamics ([T]RPMD) \cite{Craig2004, Habershon2013, Rossi2014,Rossi2018,Welsch2016} and centroid molecular dynamics (CMD) \cite{Cao1994, Cao1994a, Hone2006,Medders2015,Suleimanov2016}, have been developed for approximating nuclear quantum  effects in simulations of dynamical properties. These methods resemble classical MD in the extended phase-space of the imaginary-time Feynman paths or `ring-polymers'. They were introduced heuristically, but can also be thought of as approximations to `Matsubara dynamics' \cite{Hele2015a, Trenins2018,Jung2019a,Althorpe2021,Chowdhury2021a}, which is the classical, quantum-Boltzmann-conserving, dynamics that emerges when jagged imaginary-time paths are smoothed to remove real-time quantum coherence; [T]RPMD is a short-time approximation to Matsubara dynamics; CMD is a mean-field approximation, equivalent to averaging out the quantum fluctuations about the ring-polymer centroids \cite{Hele2015,Hele2017,Althorpe2021}.

The recently developed quasicentroid molecular dynamics (QCMD) \cite{Trenins2019} method has been applied so far only to water, and is similar to CMD, except that the fluctuations are mean-field averaged around a curvilinear `quasicentroid' rather than a cartesian centroid. The resulting QCMD infrared spectra (of gas-phase water and of the q-TIP4P/F \cite{Habershon2009} model of liquid water and ice) show none of the artificial red-shifting and broadening that affects the CMD stretch band in water (below 400 K in the gas-phase and 300 K in the liquid), with the positions of the bands lining up almost perfectly with the exact quantum spectrum (with a small temperature-independent blue shift), and the intensities of the fundamental bands approximating well the intensities of the quantum bands (in gas-phase water) \cite{Trenins2019,Benson2019}. These results are very promising, although QCMD is currently roughly 100 times more expensive than TRPMD, which gives band positions for water which are as good as QCMD but artificially broadens  the lineshapes.

The reason QCMD works so well for water is that each point on the quasicentroid potential of mean force corresponds to an ensemble of ring polymers with the same average bond length. At low temperatures, this condition prevents polymers that stretch around rotational (or librational) curves from spuriously lowering the free energy near inner turning points by forming artificial instantons, as happens in CMD \cite{Trenins2018, Witt2009, Ivanov2010}. As a result, QCMD gives more compact ring-polymer distributions than CMD, and hence a better approximation to Matsubara dynamics. 

Here, we investigate whether these advantages can be generalised to molecules other than water by extending QCMD to treat gas-phase ammonia. A (non-dissociating) water molecule is especially simple to treat because it has no internal degrees of freedom capable of highly anharmonic large-amplitude motion. We know from recent studies of overtones, combination bands and Fermi resonances, that coupling between the centroid and the fluctuation modes can have a major effect when anharmonicity is important \cite{Ple2021,Benson2021}. Such coupling is neglected by QCMD (and also by CMD, TRPMD and classical MD). We therefore pay particular attention in what follows to how well QCMD performs for the symmetric stretch band in ammonia, which correlates with the inversion pathway.

\section{Methodology}

It is straightforward to extend the QCMD treatment of water in ref.~\onlinecite{Trenins2019} to ammonia. We define curvilinear centroids
\begin{align}
    R_i = \frac{1}{N}\sum_{k=1}^N r_i^{(k)} \;\;\;\;\;\; 
    \Theta_i = \frac{1}{N}\sum_{k=1}^N \theta_i^{(k)}
\end{align}
where $i=1,2,3$ denotes the bond angle or bond length, and $k=1,\dots,N$ denotes the imaginary-time replica or `bead'. For each replica, the bond lengths and angles are functions of the cartesian bead coordinates of the four atoms, which we will refer to collectively as ${\bf q}\equiv\{q^{(k)}_l\}$, with $k=1,\dots,N$ and $l=1,\dots,12$.
These coordinates roughly correspond to the stretch and bend vibrational normal mode coordinates. In addition to these 6 internal coordinates (DOFs), 6 external coordinates are needed to uniquely define the quasicentroid configuration. These are obtained from Eckart constraints \cite{Eckart1935, Jorgensen1978, Kudin2005} similar to those in ref.~\onlinecite{Trenins2019}, which orient the ring-polymer with respect to the quasicentroid unit. We will denote the full set of quasicentroid coordinates ($\{R_i,\Theta_i\}$ plus the Eckart constraints) by $\boldsymbol{\xi}$.

The QCMD potential of mean force ${\cal F}(\boldsymbol{\xi})$ is defined to be the free energy obtained by averaging the ring-polymer fluctuations around $\boldsymbol{\xi}$, namely
\begin{equation}\label{Zqc}
e^{-\beta{\cal F}(\boldsymbol{\xi})} = Z_{qc}(\boldsymbol{\xi}) = \int\! d\textbf{p}' \int\!d\textbf{q}'\, e^{-\beta_N W_N(\textbf{q}', \textbf{p}')} \;\delta(\boldsymbol{\xi}' -\boldsymbol{\xi})
\end{equation}
where $W_N(\textbf{q},\textbf{p})$ is the ring-polymer Hamiltonian, and ${\bf p}'$ are the cartesian momenta conjugate to the cartesian bead coordinates ${\bf q}'$.
The mean forces and torques on the quasicentroid are thus
\begin{align}\label{QCcurvforce}
    f_{R_i}(\boldsymbol{\xi}) &= -  \frac{\partial {\cal F}(\boldsymbol{\xi})}{\partial R_i}  \approx -\frac{1}{N} \left \langle  \sum_{k=1}^N \frac{\partial U_N({\bf q})}{\partial r_i^{(k)}} \right \rangle_{\boldsymbol{\xi}} \\
    \label{qth} f_{\Theta_i}(\boldsymbol{\xi}) &= -\frac{\partial {\cal F}(\boldsymbol{\xi})}{\partial \Theta_i}  \approx -\frac{1}{N}
    \left \langle
    \sum_{k=1}^N \frac{\partial U_N({\bf q})}{\partial \theta_i^{(k)}}\right \rangle_{\boldsymbol{\xi}}
\end{align}
where  $U_N({\bf q})$ is the ring-polymer potential energy minus the spring term, and
\begin{equation}\label{meanfieldavg}
    \left \langle \dots \right \rangle_{{\boldsymbol{\xi}}} = \frac{1}{Z_{qc}(\boldsymbol{\xi})} \int\! d\textbf{p}' \int\!d\textbf{q}'\, e^{-\beta_N W_N(\textbf{q}', \textbf{p}')} \dots \;\delta(\boldsymbol{\xi}' -\boldsymbol{\xi})
\end{equation}
The approximations in \eqnn{QCcurvforce}{qth} follow from assuming that the ring-polymer distribution is sufficiently compact that the polymer-spring contribution to the force is negligible. As discussed in ref.~\onlinecite{Trenins2019}, this assumption of compactness also allows one to propagate the quasicentroid dynamics in cartesian coordinates (so that one does not need to work with curvilinear momenta and mass matrices), and (most crucially) to use the dynamics of the quasicentroid as a proxy for the dynamics of the centroid. Each of these approximations contributes an error to the static quantum Boltzmann distribution (sampled by the quasicentroid dynamics), and it is thus necessary to check that the QCMD static properties agree well with those of standard PIMD.

\section{Numerical details}

To propagate the QCMD dynamics of (gas-phase) ammonia, we converted the curvilinear forces of \eqnn{QCcurvforce}{qth} to cartesians (using the relations given in the Supplementary Material), and propagated the dynamics using the  adiabatic  QCMD (AQCMD) algorithm of Trenins \textit{et al.} \cite{Trenins2019} at 300 K (using $N=32$ replicas) and 150~K ($N=64$), using the ammonia potential energy and dipole-moment surfaces of Yurchenko \textit{et al.} \cite{Yurchenko2009}
At each temperature, a total of 32 initial geometries were equilibrated for 15 ps using TRPMD with a global path-integral Langevin thermostat (PILE-G) \cite{Ceriotti2010} ($\lambda = 0.5,\ \tau_ 0 = 100$ fs), after which they followed QCMD dynamics for 2 ps (to re-equilibrate to the QCMD distribution), then QCMD for a further 5 ps (production runs). The AQCMD algorithm used a timestep of $0.1 / \gamma$ fs with $\gamma=32$ (300 K) and 64 (150 K),  with a local path-integral Langevin thermostat (PILE-L) attached to the ring-polymers and a global Langevin thermostat \cite{BUSSI200826} to the quasicentroids, as in ref.~\onlinecite{Trenins2019}.

During QCMD simulations, we observed (rare) numerical instabilities occurring in the vicinity of inversion events.\footnote{The inversion process is a thermally activated rare event along the symmetric bend coordinate, promoted primarily by ring polymer fluctuations. In QCMD, these fluctuations do not sufficiently carry over to the quasicentroid due to strong anharmonicity, so, beyond the harmonic limit the quasicentroid is no longer close enough to the centroid and the QCMD approximation breaks down.} A small fraction of the QCMD trajectories ($\sim1.6 \% $) were hence discarded when the ammonia attempted to invert, since the Eckart constraint cannot handle large amplitude motion (LAM) \cite{Szalay1988, Sayvetz1939} \footnote{Here we are referring to the LAM of the centroid from the quasicentroid, not the inversion pathway (which is itself a LAM).}.

\begin{figure}
    \includegraphics[width=\textwidth]{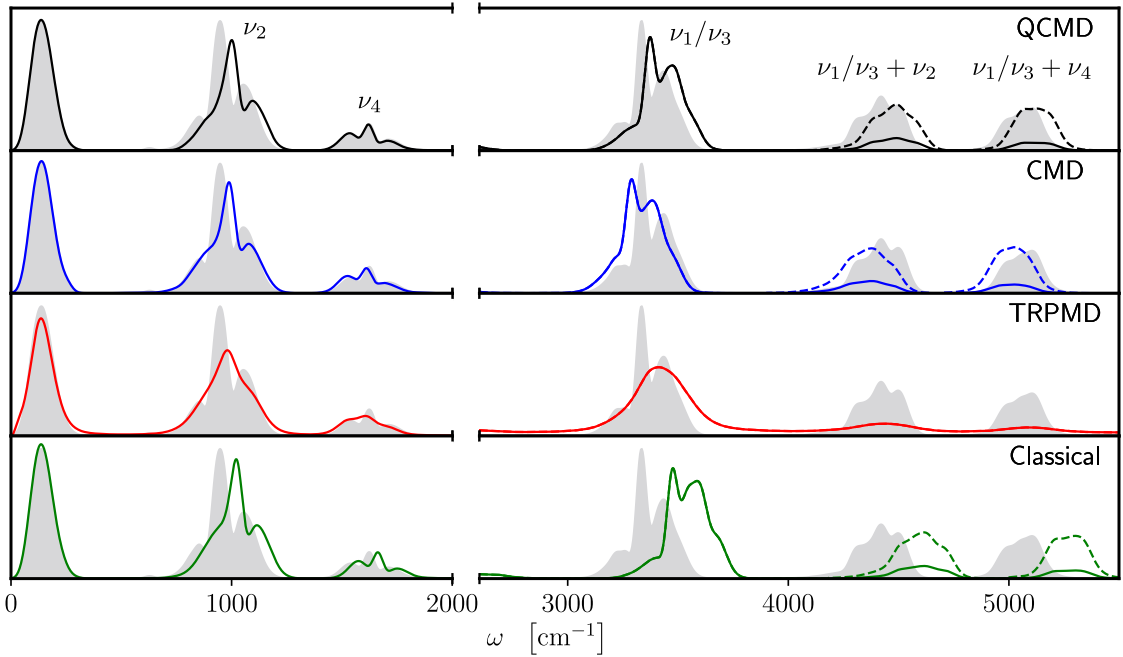}
    \caption{\label{ammonia300K} Infrared absorption spectra for gas-phase ammonia at 300 K, computed by extending QCMD as described in the text (black line), compared with the damped quantum spectrum adapted from ref.~\onlinecite{Yurchenko2009} (grey shading). Also shown are the results of standard CMD,  TRPMD and classical MD simulations. The dashed lines were obtained by multiplying the $\nu_1/\nu_3 + \nu_2$ and $\nu_1/\nu_3 + \nu_4$ combination bands (except for the TRPMD bands,  which have very wide tails) by the post-processing correction factor of ref.~\onlinecite{Ple2021,Benson2021}. The absorption intensities in the two panels are scaled in the ratio 1:13.}
\end{figure}

\begin{figure}
    \includegraphics[width=\textwidth]{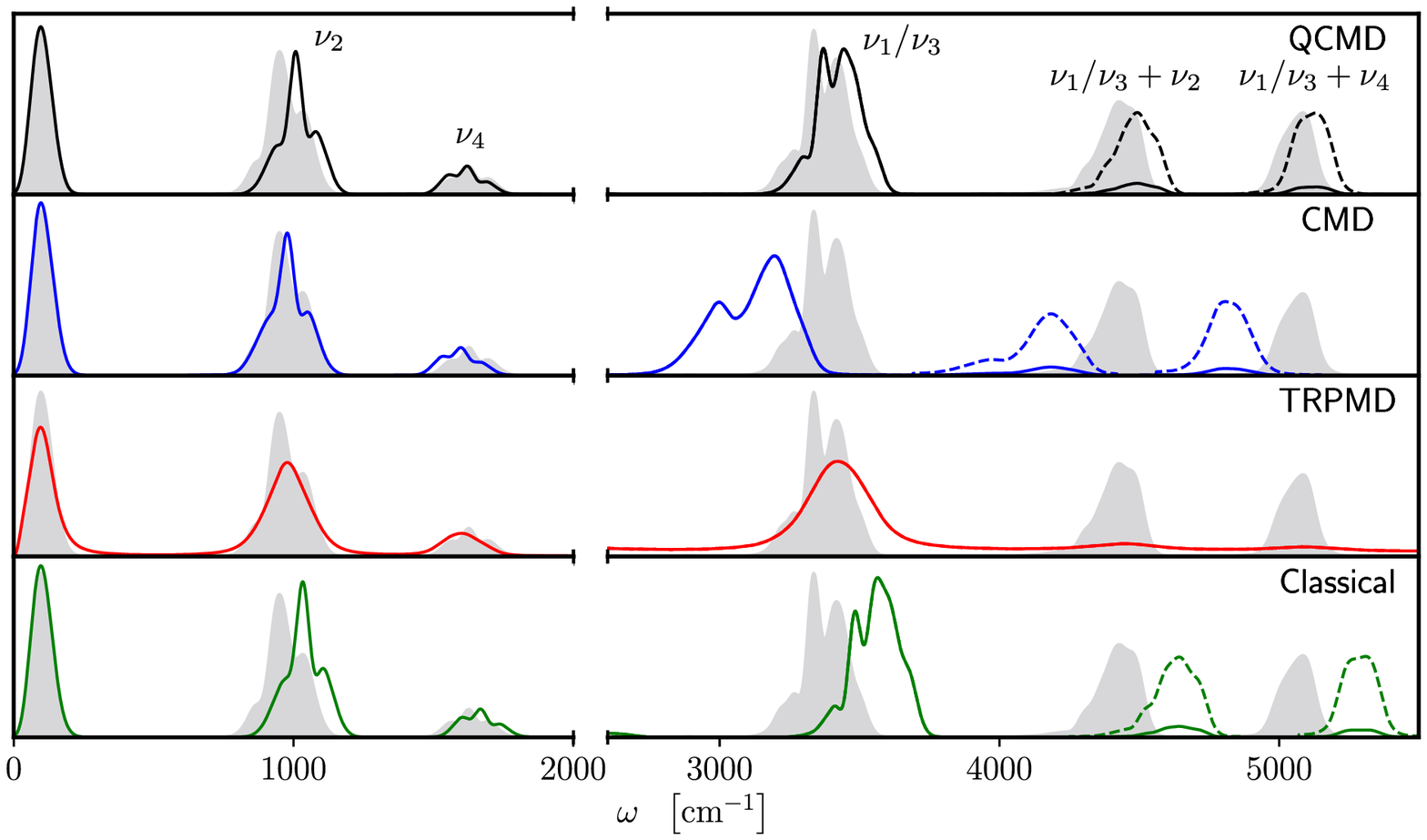}
    \caption{Same as Fig.~\ref{ammonia300K} for 150 K. The absorption intensities in the two panels are scaled in the ratio 1:19.}
    \label{ammonia150K}
\end{figure}

The spectra were obtained from the dipole derivative time-autocorrelation function $C_{\mathbf{\dot{\mu}}\cdot\mathbf{\dot{\mu}}}(t)$. 
Since volume is not defined for the gas-phase simulations, the power spectrum
\begin{equation}
	\tilde{I}_{\mathbf{\dot{\mu}}} \left(\omega\right) = \beta\int_{-\infty}^\infty\!dt\ e^{-i\omega t} C_{\mathbf{\dot{\mu}}\cdot\mathbf{\dot{\mu}}}(t)f(t)
\end{equation}
was calculated in lieu of the IR spectrum \cite{Habershon2008}, where $f(t)$ is a Hann window of width  $\tau = 750$ fs.

\section{Results}

The resulting QCMD spectra are shown in Figs.~1 (300 K) and 2 (150 K), where they are compared with the results of CMD and TRPMD  (computed using standard PIMD methods), classical MD, and quantum dynamics (obtained by Boltzmann-weighting the line list found in ref.~\onlinecite{Yurchenko2009} and convolving with $f(t)$). The combination-band regions are also shown multiplied by the perturbative correction factor of ref.~\onlinecite{Yao1976}, which was found in refs.~\onlinecite{Benson2021} and \onlinecite{Ple2021} to account for most of the increase in intensity of the non-fundamental bands that results from the neglected coupling between the Matsubara dynamics of the centroid and the fluctuation modes.

With the exception of the symmetric bend ($\nu_2$), QCMD performs almost as well for ammonia as for gas-phase water. The QCMD asymmetric bend ($\nu_4$) and stretch bands ($\nu_1/\nu_3$) line up with the TRPMD bands, but are not artificially broadened. Both QCMD and TRPMD give a similar (temperature-independent) blue shift for the stretch band for ammonia (35 cm$^{-1}$) to gas-phase water (60 cm$^{-1}$), because they are both affected by the neglect of centroid-fluctuation coupling and real-time coherence. The QCMD combination bands are also close in position to the quantum bands and the intensities are in good agreement when multiplied by the perturbative correction factor (dashed lines in Figs.~1 and 2). This suggests that the ring-polymer distributions around the quasi-centroids are compact, such that QCMD gives a good approximation to Matsubara dynamics, and also samples a good approximation to the exact quantum Boltzmann statistics. This is borne out by the QCMD average bond lengths and bond angles, which are within better than $0.2\%$ of the values computed using standard PIMD.

\begin{figure}
    \includegraphics[scale=0.5]{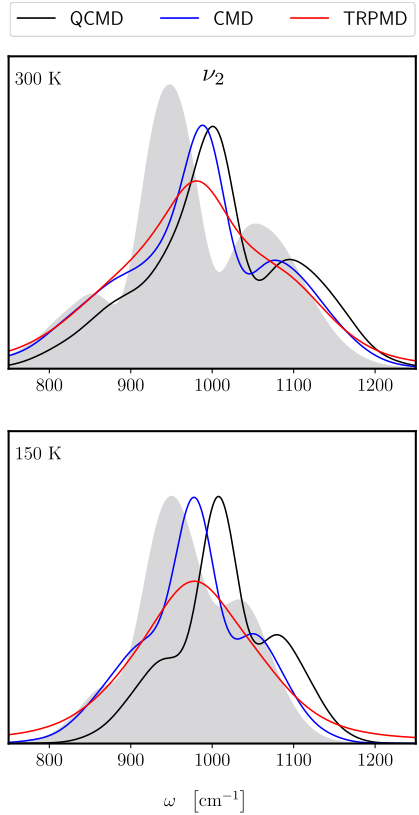}
    \caption{Expanded plot of the symmetric-bend bands of Figs.~1 and 2.}
    \label{bend mode}
\end{figure}

\begin{figure}
    \centering
    \includegraphics[scale =0.8]{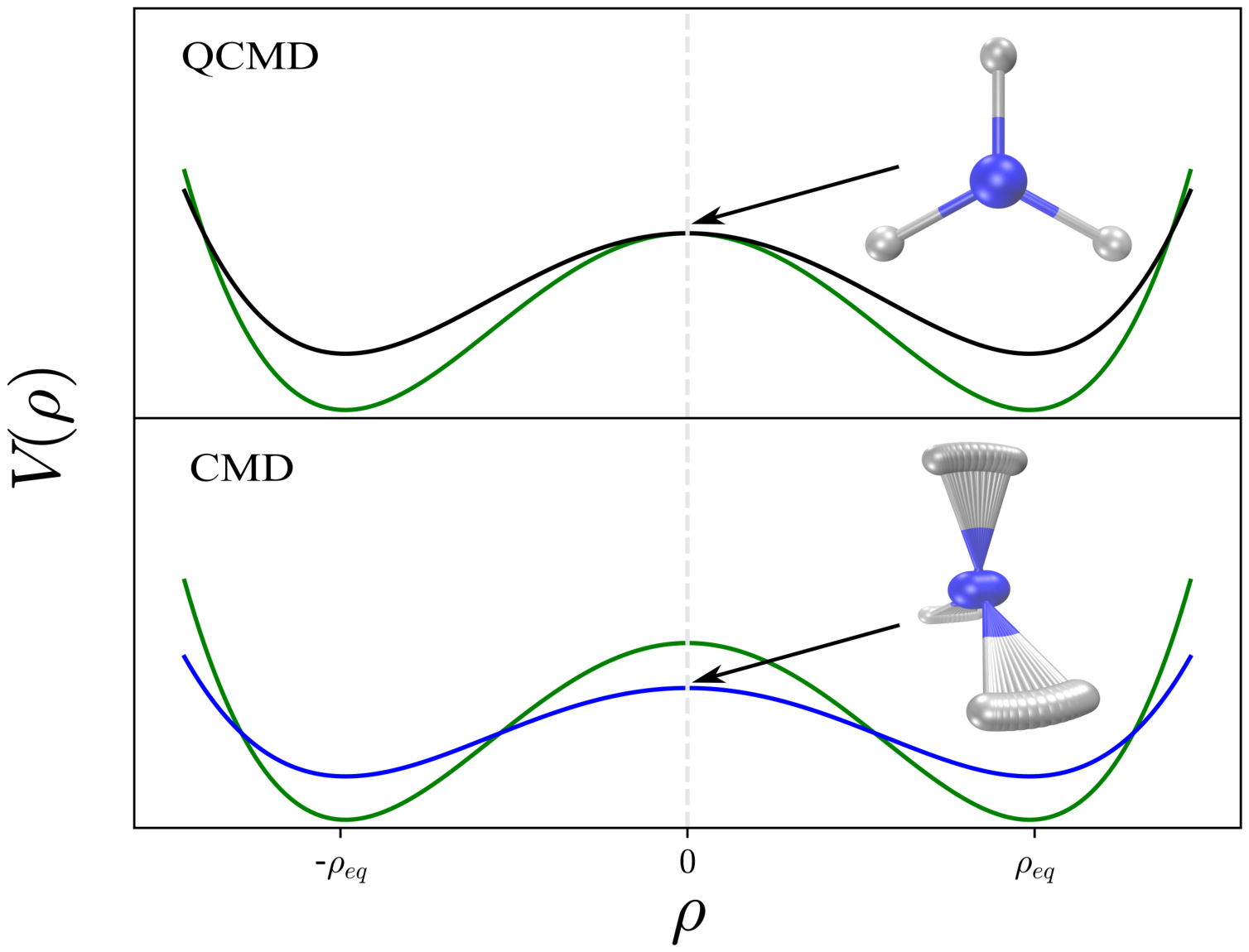}
    \caption{Schematic plot of the potential energy (green) and the  potentials of mean force for QCMD (black) and CMD (blue) along the inversion coordinate $\rho$.  The QCMD curvilinear centroid constraint artificially prevents the formation of the tunnelling instanton at the barrier.}
    \label{Inversion potential}
\end{figure}

For the symmetric bend, QCMD gives a $\sim$50~cm$^{-1}$ temperature-dependent blue shift with respect to the quantum band, making
this the only QCMD band that does not line up with TRPMD at 150 K (see Fig.~3). Discarding the inverting QCMD trajectories (see above) \footnote{The Eckart conditions fail to invert the quasicentroid unit when the ring-polymer inverts, which necessitated the discarding of inverting trajectories. Since the timescale of this inversion process is higher than any of the vibration timescales, this procedure ought to have a minimal effect on the spectrum. } may have biased the sampling of this band and thus skewed its shape, but such an error is likely to be minor, since discarding the inverting trajectories in TRPMD does not affect its spectrum. There may also be an error in this band from the inability of QCMD and TRPMD to describe coherent tunnelling and thus to reproduce the 35~cm$^{-1}$ bend-tunnelling splitting \cite{Yurchenko2005}, although such an error would be unlikely to shift the overall position of the band, and presumably has a similar effect on both QCMD and TRPMD. Nevertheless, the (planar) inversion barrier does give a clue as to what causes the QCMD blue shift. In CMD (and TRPMD), the ring-polymer lowers its potential energy at the barrier by forming a (non-artificial) instanton \cite{Richardson2009} along the inversion coordinate $\rho$ \cite{Yurchenko2009},
\begin{align}
   \rho = \arccos\frac{2}{\sqrt{3}} \sin { \left(\frac{ \theta_1+\theta_2+\theta_3}{6}\right)}
\end{align}
as illustrated in Fig.~4 \footnote{Note that $\rho$ defined here and the inversion coordinate $\overline{\rho}$ in ref.\citenum{Yurchenko2009} are related as $\rho = \frac{\pi}{2} - \overline{\rho}$}. However, in QCMD, the curvilinear centroid constraint at the barrier ($\Theta_1+\Theta_2+\Theta_3=2\pi$) forces each of the $N$ ammonia replicas to be planar, which prevents the instanton from forming.\footnote{Strictly, the quasicentroid constraint does not prevent the instanton geometry from forming, but assigns it to a quasicentroid ensemble that corresponds to a bent geometry, where its contribution to the free energy is negligible. } Although the barrier itself is not important, we can expect that a similar constriction in the range of geometries sampled also increases the QCMD potential of mean force on the barrier-side of the potential wells. 

In other words, the QCMD curvilinear centroid constraints, which annihilate the artificial instantons,  also annihilate the {\em genuine} instanton at the barrier top, thus giving an artificial blue shift in the symmetric bend. The CMD cartesian centroid constraints permit both the artificial and the genuine instantons to form, which seems to have resulted in an almost perfect cancellation of errors in the frequency of the symmetric bend  at 150 K (but not at other temperatures).
Of the methods tested, the best estimate of the symmetric bend frequency is given by TRPMD, which does not interfere with the instanton, and thus exerts the correct force on the centroid (although not on the fluctuation modes). Note that a variety of other semiclassical methods, including path-integral Liouville dynamics\cite{Liu2016} and semi-classical initial value representation (SC-IVR) methods\cite{Miller2001,Conte2013}, also give a good estimate of this frequency.

It is tempting to try to fix this problem by designing a centroid coordinate that annihilates artificial instantons but preserves real ones. However, besides being messy and semi-empirical, such a proposed solution neglects the real problem here, which is that the anharmonic 
Matsubara dynamics of the symmetric bend involves so much centroid-fluctuation coupling that is probably necessary to remove at least some of the fluctuation modes from the mean-field, and to explicitly couple their dynamics to that of the quasicentroid. Such an approach would re-introduce the Matsubara phase, and is thus unlikely to be practical.

However, we should keep in mind that the `bad' symmetric-bend frequency is no worse than that predicted by classical MD, and that  overall the agreement between QCMD and the quantum spectrum (Figs.~1 and 2) is a big improvement on that provided by TRPMD and CMD.

\section{Conclusions}

These tests on gas-phase ammonia suggest that the advantages of QCMD can be generalised to many systems other than water, 
with the important caveat that degrees of freedom that are both 
(Boltzmann-statistically) quantum and highly anharmonic (such as the ammonia inversion mode) are unlikely to be treated better by QCMD than by other path-integral methods. 

We expect these findings to transfer straightforwardly to the condensed phase, since intermolecular degrees of freedom (such as librations and centre-of-mass vibrations) tend to be Boltzmann-statistically classical. QCMD has in fact already been applied to the q-TIP4P/F model of liquid water and ice\cite{Trenins2019}. Similar calculations would be straightforward for liquid ammonia (since the only modification needed to Sec.~II would be the addition of the torque estimator $\boldsymbol{\tau}$, as discussed in ref.~\onlinecite{Trenins2019}). More generally, applications of QCMD to mixtures of water and organic molecules, e.g.\ clathrates or water-organic solvent interfaces should be possible. However, QCMD is unlikely
to be able to treat proton-transfer processes (in solution or in the gas-phase), since proton-transfer coordinates are typically (Boltzmann-statistically) quantum and highly anharmonic\cite{Yu2019}.

The `elephant in the room' is that QCMD is  typically 10-100 times more expensive than competing path-integral dynamics methods (such as TRPMD and CMD),  on account of the adiabatic algorithm used to generate the quasicentroid
potential of mean force. More efficient algorithms will need to be developed if QCMD is to become widely applicable.

\section*{Supplementary Material}

See the supplementary material for further details of the cartesian to curvilinear coordinate transformations.

\begin{acknowledgments}
It is a pleasure to thank Venkat Kapil for insightful discussions about path-integral simulation.
C.H. acknowledges support from the EPSRC Centre for Doctoral Training in Computational Methods for Materials Science (Grant No. EP/L015552/1); V.G.S.\ from St.\ John's College, University of Cambridge, through a Dr Manmohan Singh Scholarship;  G.T.\ from the Cambridge Philosophical Society and St Catharine's College, University of Cambridge. 
\end{acknowledgments}

\section*{Conflicts of Interest}

The authors have no conflicts to disclose.

\section*{Author Contributions}

C.H. and V.G.S. contributed equally to this work.

\section*{Data Availability}

The data that supports the findings of this study are available within the article.

\section*{Credit Line}

This article may be downloaded for personal use only. Any other use requires prior permission of the author and AIP Publishing. This article appeared in C. Haggard, V. Sadhasivam, G. Trenins and S. Althorpe, J. Chem. Phys., \textbf{155}, 174120 (2021). and may be found at \href{https://doi.org/10.1063/5.0068250}{\color{blue}https://doi.org/10.1063/5.0068250}.

\selectlanguage{english}
\bibliography{Citations}

\end{document}

\date{}
\maketitle

\section*{Transformation of forces:}
Let $(r_1, r_2, r_3, \theta_1, \theta_2, \theta_3)$ denote the bond lengths and bond angles respectively in ammonia, as shown in the figure below:

\begin{figure}[H]
    \centering
    \includegraphics[scale=0.25]{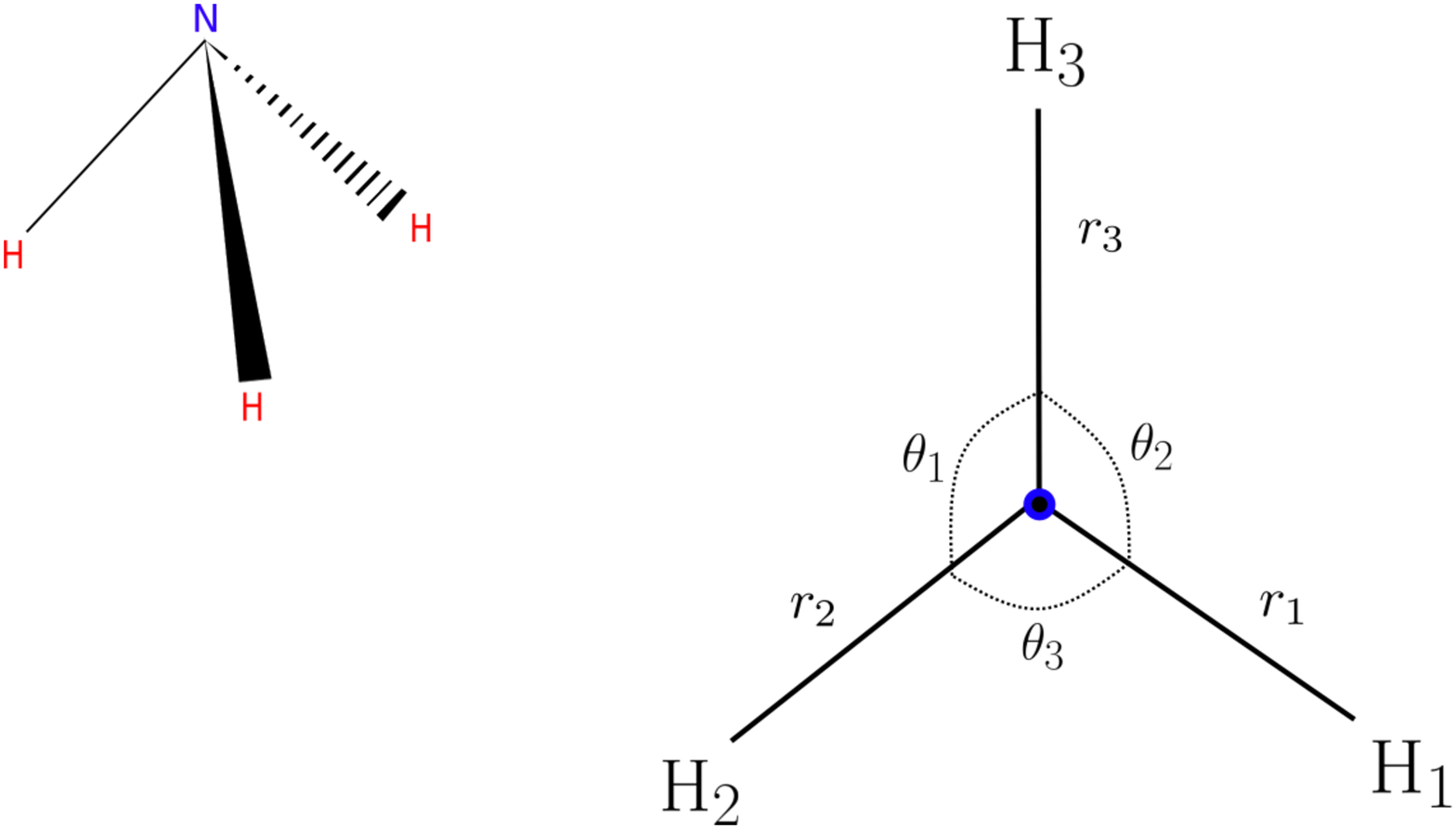}
    \caption{Bond angles and bond lengths in ammonia}
    \label{NH3}
\end{figure}

We assume that the potential energy $U(\textbf{Q})$ of an ammonia molecule can be completely described in terms of these curvilinear coordinates. (Note here that $\textbf{Q}$ denotes a general point in space in Cartesian coordinates, not necessarily the quasicentroid or bead position). Then, for a given configuration of atoms in a molecule denoted as $\textbf{Q}_{H_1}$, $\textbf{Q}_{H_2}$, $\textbf{Q}_{H_3}$ and $\textbf{Q}_{N}$, the cartesian forces on the atoms are given by:

\begin{align}
\textbf{f}_{\alpha} = -\frac{\partial U(\textbf{Q})}{\partial \textbf{Q}_{\alpha} } &= - \left( \sum_{i=1}^{3} \frac{\partial U(\textbf{Q})}{\partial r_i } \nabla_{\textbf{Q}_{\alpha}} (r_i) + \sum_{i=1}^{3} \frac{\partial U(\textbf{Q})}{\partial \theta_i } \nabla_{\textbf{Q}_{\alpha}} (\theta_i) \right) \\
\Longrightarrow f_{\alpha} &=  \sum_{i=1}^{3} f_{r_i} \nabla_{\textbf{Q}_{\alpha}} (r_i) + \sum_{i=1}^{3}  f_{\theta_i}  \nabla_{\textbf{Q}_{\alpha}} (\theta_i) 
\end{align}
where $\alpha = (\text{H}_1, \text{H}_2, \text{H}_3, \text{N})$. 

We derive the cartesian force on one of the hydrogen atoms, say $\text{H}_1$. The forces on the other two can be obtained by exploiting the $C_{3v}$ symmetry of ammonia. The force on the nitrogen atom can be obtained as:

\begin{align}\label{nonetforceeq}
    \textbf{f}_{_N} = -(\textbf{f}_{{H_1}} + \textbf{f}_{{H_2}} +  \textbf{f}_{{H_3}})
\end{align}
as the total force on the molecule adds up to zero. 

We begin by considering the transformation from cartesian to curvilinear coordinates as follows:

\begin{align}
    r_i = \norm{\textbf{Q}_{NH_i}} = \norm{\textbf{Q}_{H_i} - \textbf{Q}_N} \\
    \theta_i = \text{cos}^{-1} \left(\frac{\textbf{Q}_{NH_j}\vdot{\textbf{Q}_{NH_k}}}{r_j \;r_k}\right)
\end{align}

From the above expressions, it is clear that $ \nabla_{\textbf{Q}_{H_1}} (r_2)$, $ \nabla_{\textbf{Q}_{H_1}} (r_3)$ and $ \nabla_{\textbf{Q}_{H_1}} (\theta_1) $ vanish thereby yielding:

\begin{align}
    \frac{\partial U(\textbf{Q})}{\partial \textbf{Q}_{H_1}} = \frac{\partial U}{\partial r_1} \nabla_{\textbf{Q}_{H_1}} (r_1) + \frac{\partial U}{\partial \theta_2} \nabla_{\textbf{Q}_{H_1}} (\theta_2) +   \frac{\partial U}{\partial \theta_3} \nabla_{\textbf{Q}_{H_1}} (\theta_3) 
\end{align}

It is easy to check that:
\begin{align}
    \nabla_{\textbf{Q}_{H_1}} (r_1) &= \boldsymbol{\xi}_1 \;\;\;\;\;  \\\nabla_{\textbf{Q}_{H_1}} (\theta_2) = \boldsymbol{\Phi}_{213} \;\;\;\; &\text{and} \;\;\;\; \nabla_{\textbf{Q}_{H_1}} (\theta_{3}) = \boldsymbol{\Phi}_{312}
\end{align}

where 
\begin{align}
&\boldsymbol{\xi}_i := \frac{\textbf{Q}_{H_i}}{r_i} \\
    \boldsymbol{\Phi}_{ijk} := &\frac{1}{r_j \; \sin\theta_i} \left(  \frac{\textbf{Q}_{NH_j}}{r_j}\cos\theta_i  - \frac{\textbf{Q}_{NH_k}}{r_k} \right)
\end{align}

Thus, we have:
\begin{align}\label{fH1}
    \textbf{f}_{{H_1}} = f_{r_1}\boldsymbol{\xi}_1+ f_{\theta_2} \boldsymbol{\Phi}_{213} + 
    f_{\theta_3} \boldsymbol{\Phi}_{312}
\end{align}

The forces on $H_2$ and $H_3$ can be obtained by cyclically permuting the indices in eq. \eqref{fH1} , yielding:

\begin{subequations}\label{Cartesianforces}
\begin{align}
\textbf{f}_{{H_1}} = f_{r_1}\boldsymbol{\xi}_1+ f_{\theta_2} \boldsymbol{\Phi}_{213} + 
    f_{\theta_3} \boldsymbol{\Phi}_{312} \\
\textbf{f}_{{H_2}} = f_{r_2}\boldsymbol{\xi}_2+ f_{\theta_3} \boldsymbol{\Phi}_{321} + 
    f_{\theta_1} \boldsymbol{\Phi}_{123} \\
\textbf{f}_{{H_3}} = f_{r_3}\boldsymbol{\xi}_3+ f_{\theta_1} \boldsymbol{\Phi}_{132} + 
    f_{\theta_2} \boldsymbol{\Phi}_{231}
\end{align}
\end{subequations}

Now, in order to obtain the curvilinear forces, given the cartesian forces on the atoms, we start from eq. \eqref{Cartesianforces} and take a dot product with the auxiliary variables $\boldsymbol{\xi}_i$ and $\boldsymbol{\Phi}_{ijk}$. It can be easily seen that:
\begin{align}\label{curvforceR}
     f_{r_i} = \textbf{f}_{{H_i}} \vdot \boldsymbol{\xi}_i
\end{align}

The derivation of the forces along the bond angles is slightly more involved. Considering one of the hydrogen atoms, say $H_1$ as before, we get:
\begin{subequations}\label{thetaeqns}
\begin{align}
    \textbf{f}_{{H_1}} \vdot \boldsymbol{\Phi}_{213} = \frac{1}{r_1^2}f_{\theta_2} +  \frac{\zeta_{123}}{r_1^2}f_{\theta_3} \\
    \textbf{f}_{{H_1}} \vdot \boldsymbol{\Phi}_{312} =  \frac{\zeta_{123}}{r_1^2}f_{\theta_2} + \frac{1}{r_1^2}f_{\theta_3}  
\end{align}
\end{subequations}
where
\begin{align}
    \zeta_{ijk} = \frac{\cos\theta_i - \cos\theta_j\cos\theta_k }{\sin\theta_j\sin\theta_k}
\end{align}

Solving the linear equations \eqref{thetaeqns} yields:
\begin{align}\label{forceth2}
    f_{\theta_2} = \frac{r_1^2}{\zeta_{123}^2 - 1} \left( \zeta_{123} \boldsymbol{\Phi}_{312} - \boldsymbol{\Phi}_{213} \right) \vdot \textbf{f}_{{H_1}} \\\label{forceth3}
    f_{\theta_3} = \frac{r_1^2}{\zeta_{123}^2 - 1} \left( \zeta_{132} \boldsymbol{\Phi}_{213} - \boldsymbol{\Phi}_{312} \right) \vdot \textbf{f}_{{H_1}}
\end{align}

The force along $\theta_1$ can be obtained by cyclically permuting the indices in one of eqs. \eqref{forceth2} and \eqref{forceth3}. The final equations for the curvilinear forces are as follows: 

\begin{subequations}\label{curvforceTheta}
\begin{align}
    f_{\theta_1} = \frac{r_2^2}{\zeta_{213}^2 - 1} \; \boldsymbol{\chi}_{213}\vdot \textbf{f}_{{H_2}}  \\
    f_{\theta_2} = \frac{r_3^2}{\zeta_{321}^2 - 1} \; \boldsymbol{\chi}_{321}\vdot \textbf{f}_{{H_3}} \\
    f_{\theta_3} = \frac{r_1^2}{\zeta_{132}^2 - 1} \; \boldsymbol{\chi}_{132}\vdot \textbf{f}_{{H_1}}
\end{align}
\end{subequations}
where 
\begin{align}
    \boldsymbol{\chi}_{ijk} = \zeta_{ijk}\boldsymbol{\Phi}_{kij} - \boldsymbol{\Phi}_{jik}
\end{align}

Note that several equivalent expressions can be derived for the curvilinear forces. 

\section*{Derivation of cartesian forces on the quasicentroids:}
The curvilinear centroids for ammonia (or any tetra-atomic system with a single heavy atom attached to three lighter atoms) can be defined as the mean of the bond lengths and bond angles of the ring polymer beads, as mentioned in eq. 3 and 4 of the main text. For an $N$-bead ring polymer, these are:

\begin{align}
    R_i = \frac{1}{N}\sum_{k=1}^N r_i^{(k)} \;\;\;\;\;\; 
    \Theta_i = \frac{1}{N}\sum_{k=1}^N \theta_i^{(k)}
\end{align}
where $r_i^{(k)}$ and $\theta_i^{(k)}$ are the $i$-th bond length and bond angle respectively of the $k$-th bead. 

It follows that the force on the quasicentroid in curvilinear coordinates $(R_1,R_2,R_3,\Theta_1,\Theta_2,\Theta_3)$ is given by:
\begin{align}\label{QCcurvforce}
    f_{R_i} = -\frac{\partial U_N(\textbf{q})}{\partial R_i} = -\frac{1}{N} \sum_{k=1}^N \frac{\partial U_N(\textbf{q})}{\partial r_i^{(k)}} \;\;\;\;\;\;\;\;\;\;\;\; f_{\Theta_i} = -\frac{\partial U_N(\textbf{q})}{\partial \Theta_i} = -\frac{1}{N} \sum_{k=1}^N \frac{\partial U_N(\textbf{q})}{\partial \theta_i^{(k)}}
\end{align}
where $U_N(\textbf{q}) = \sum_{i=1}^N U(\textbf{q}_i)$ as defined in eq. 3 and 4 in the main text. 

From the potential energy surface, we get the cartesian forces on the beads, which can be transformed to curvilinear coordinates using eq. \eqref{curvforceR} and \eqref{curvforceTheta}. This gives the force on the quasicentroids in curvilinear coordinates, from which cartesian quasicentroid forces can be obtained using \eqref{Cartesianforces} and \eqref{nonetforceeq}.